\documentclass[aps,prl,twocolumn,showpacs,preprintnumbers,amsmath,amssymb]{revtex4}
\input epsf  \usepackage{graphicx} \usepackage{dcolumn} \usepackage{bm}

\begin{document}

\title{Microscopic mechanism for cold denaturation}

\author{Cristiano L. Dias$^{1}$, Tapio Ala-Nissila$^{2,3}$, Mikko
Karttunen$^{4}$, Ilpo Vattulainen$^{5,6,7}$ and Martin Grant$^{1}$ }
\affiliation{ $^{1}$Physics Department, Rutherford Building, McGill
University, 3600 rue University, Montr\a'eal, Qu\a'ebec,  H3A 2T8
Canada\\ $^{2}$ Department of Physics, Brown University, Providence RI
02912-1843\\ $^{3}$ Laboratory of Physics, Helsinki University of
Technology, P.O. Box 1100, FI--02015 TKK, Espoo, Finland\\ $^{4}$
Department of Applied Mathematics, The University of  Western Ontario,
London, Ontario, Canada\\  $^{5}$ Institute of Physics, Tampere
University of Technology  P.O. Box 692, FI--33101 Tampere, Finland\\
$^{6}$ MEMPHYS--Center for Biomembrane Physics,  University of
Southern Denmark\\ $^{7}$ Helsinki Institute of Physics, Helsinki
University of Technology, Finland}

\begin{abstract}
We elucidate the mechanism of cold denaturation through
constant-pressure simulations for a model of hydrophobic molecules  in
an explicit solvent. We find that the temperature dependence of the
hydrophobic effect is the driving force/induces/facilitates cold
denaturation. The physical mechanism underlying this phenomenon is
identified as the destabilization of hydrophobic contact in favor of
solvent  separated configurations, the same mechanism seen in pressure
induced denaturation. A phenomenological explanation proposed for  the
mechanism is suggested as being responsible for cold  denaturation in
real proteins.
\end{abstract}


\maketitle

Under physiological conditions, proteins adopt a unique
three-dimensional (3D) structure \cite{ANFI73}. It is maximally stable
at about $17^{\circ}$C and becomes unstable thus denaturing the
protein  at both high ($\sim 60^{\circ}$C) and low ($\sim
-20^{\circ}$C)  temperatures \cite{PRIV86,PRIV97,KUNU02}. The latter
phenomenon  is called {\em cold denaturation}, where the protein
unfolds and  thereby increases its entropy, which in turn is
accompanied by  a decrease in the entropy of the entire system. This
counter-intuitive  behavior has been experimentally verified
\cite{PRIV97,RAVI03} but  has remained a subject of controversy
\cite{PRIV86,KUNU02}, since a  satisfactory microscopic explanation
for this phenomenon has  not yet emerged. Resolving cold denaturation
microscopically would facilitate understanding the forces responsible
for the structure of  proteins, and in particular the role of the
complex hydrophobic effect

In the case of diluted proteins, hydrophobicity is considered as  the
main driving force for folding and unfolding \cite{DILL90}.
Consequently, different classes of models describing  hydrophobicity
through varying explicit models have been used to  study cold
denaturation \cite{BRUS00,PASC05,RIOS00,COLL01}.  One such class
\cite{BRUS00,RIOS00} considers the different energetic states of shell
water, i.e. water molecules neighboring the protein, in a lattice. A
more realistic water model \cite{PASC05} supports this view, as
water-water hydrogen bonding among shell water  has been found to
increase at low temperatures and to correlate  with cold
denaturation. Meanwhile, another class of models suggests  that the
density fluctuations of water are responsible for cold  denaturation
\cite{MARQ03,BULD07}. Despite the lack of consensus in  the
explanation of cold denaturation, the solvent is widely accepted  as
the key player. This is also supported by the fact that
denaturation also takes place under pressure \cite{HUMM98,MEER06}.  By
focusing on the transfer of water molecules to the protein interior,
pressure denaturation has been explained through the destabilization
of hydrophobic contacts in favor of solvent separated configurations
\cite{HUMM98}. This destabilization has been verified using  different
water models \cite{GHOS01}.

In the present work, we  examine/unravel/reveal the microscopic
physical mechanism behind cold denaturation. To this end, we consider
the two-dimensional Mercedes-Benz (MB) model to describe water
molecules in the solvent and a simple bead-spring model for the
protein. The MB model reproduces many of the properties of water
\cite{SILV98}, including the temperature dependent behavior of the
hydrophobic effect \cite{DILL05}. Our molecular dynamics (MD)
simulations of the MB model provide a simple microscopic picture  for
cold denaturation in terms of changes in hydration:  at low
temperatures water molecules infiltrate the  folded protein in order
to passivate the ``dangling'' water-water  hydrogen bonds (H-bonds)
found in shell water. At the same time,  hydrophobic contacts are
destabilized and an ordered layer of  water molecules forms around the
protein monomers such that they become separated by a layer of solvent
in the cold denatured state.  Hence, increasing pressure and
decreasing temperature destabilize hydrophobic contacts in favor of
similar solvent separated configurations.  We expect that this
aggravated destabilization of hydrophobic contacts  at high pressure
explains why the transition temperature  for cold denaturation
increases with increasing pressure \cite{KUNU02}.  Here, we study cold
denaturation at the equivalent of ambient pressure.

As in water, the interaction between the MB molecules is given by  
a sum of hydrogen bonds and van der Waals bonds. The directionality  of
H-bonds is accounted for by three arms separated by an angle of
$120^{\circ}$. This interaction has maximal strength when arms of
neighboring molecules are aligned. If $\vec{r}_{ij}$ is the distance
vector between the center of mass of molecules $i$ and $j$, and
$\vec{r}_{i\alpha}$ is the distance vector between the center of
molecule $i$ and the extremity of arm $\alpha$, then the interaction
energy is given by:
\begin{eqnarray}
V_h (\vec{r}_{ij},\{\vec{r}_{i\alpha}\},\{\vec{r}_{j\beta}\}) =
\epsilon_h  \exp{ \left( - \frac{(r_{ij}-R_h)^2} {2\sigma_R^2}
\right)} \times \nonumber \\   \left[ \sum_{\alpha=1}^3 \exp \left( -
\left( \frac{\vec{r}_{i\alpha} \cdot \vec{r}_{ij}}{r_{ij}r_{i\alpha}}
- 1 \right)^2 \frac{1}{2\sigma_{\theta}^2} \right) \right] \times
\nonumber \\   \left[ \sum_{\beta=1}^{3} \exp \left( - \left(
\frac{\vec{r}_{j\beta} \cdot \vec{r}_{ij}}{r_{ij}r_{j\beta}}
-1\right)^2 \frac{1}{2\sigma_{\theta}^2} \right) \right] ,
\label{eqn:hydrogen_bond}
\end{eqnarray}
where $\epsilon_{h}$ and $R_h$ are the binding energy and the
equilibrium (reference) length of the bond, respectively. The
constants $\sigma_R$ and $\sigma_{\theta}$ are attenuation  parameters
of the interaction. Equation~(1) favors configurations  where the
distance between molecules $i$ and $j$ is $R_h$, one arm  of molecule
$i$ is aligned with the line joining the two centers  of mass, and the
same for one arm of molecule $j$.  The van der  Waals interaction is
described by a Lennard--Jones (LJ) potential  $V_{ww}$ with binding
energy $\epsilon_{ww}$ and equilibrium  length $R_{ww}$:
\begin{equation}
V_{ww}(r_{ij}) = 4 \epsilon_{ww} \left[ \left( \frac{R_{ww}}{r_{ij}}
\right)^{12} - \left( \frac{R_{ww}}{r_{ij}} \right)^{6}\right] .
\label{eqn:potentialww}
\end{equation}
The LJ potentials are shifted so that the force becomes zero at  the
cut-off distance $R_c = 2.5 R_h$ \cite{ALLE90}. We use the  parameter
set which has been studied extensively by Silverstein  {\em et al.}
\cite{SILV98}: $\epsilon_{h} = 1.0$, $R_{h} = 1.0$,  $\sigma_R =
\sigma_{\theta} = 0.085$, $\epsilon_{ww} = 0.1$,  and $R_{ww} =
0.7$. The total interaction energy $V_{i,j}$ between  two water
molecules is given by the sum of Eqs.~\ref{eqn:hydrogen_bond}  and
\ref{eqn:potentialww}.

Here, we set $M_w = 1$ for water. To mimic the distribution of  mass
in water, $1/10$ of the total mass of a water molecule is  located at
each arm's extremity and the extremity of an arm is  located at a
distance $R_\mathrm{arm} = 0.36 R_h$ from the center  of mass
\cite{STIL71}. This defines the angular momentum of the  water
molecule.

Energies, distances, and time are given in units of $\epsilon_h$,
$R_h$, and $\tau_o = \sqrt{\epsilon_{ww} / M_w R_{ww}^2}$,
respectively.  To model the protein, we use a bead-spring model:
monomers which  are adjacent along the backbone of the protein are
connected to  each other by springs, and non-adjacent monomers are
connected by  a shifted LJ potential. The LJ potential is described by
a binding  energy $\epsilon_{mm} = 0.375$ and distance $R_{mm}$. The
equilibrium  length and stiffness of the spring are
$R_\mathrm{spring}$ and  $K_\mathrm{spring}= 2 ( 456 \epsilon_{mm} /
R_{mm}^2 )$. This corresponds  to twice the stiffness of the LJ
potential. Monomers are set to be  ten times heavier than water
molecules. The interaction between  monomers and water molecules is
given by a shifted LJ potential with  binding energy $\epsilon_{wm} =
\epsilon_{ww}$ and equilibrium  length $R_{wm}$.

When the side-chain of a hydrophobic amino acid is exposed to the
solvent, the liquid surrounding the side-chain assumes a cage-like
configuration \cite{BOWR98} in order to minimize the amount of broken
H-bonds of water molecules. This configuration has a low entropy and
proteins minimize their free-energy by burying these hydrophobic
amino-acids in their interior. To reproduce this, we choose
$R_\mathrm{spring} = 2.0$ and $R_{wm} = 0.9$ such that monomers can be
surrounded by a layer of water molecules when exposed to the
solvent. To allow for the formation of a dry protein core, we use
$R_{mm} \backsimeq R_{wm}$, though $2 R_{mm} > R_\mathrm{spring}$  to
avoid the backbone from intersecting itself. Taking these restrictions
into account, we choose $R_{mm} = 1.1$.

Having defined the interaction between the different particles  we now
perform MD in the isothermal-isobaric ensemble. Constant  pressure is
achieved using the Andersen extended method \cite{ANDE80}.  To
suppress oscillations of the simulation box, the canonical  equations
of motion are replaced by a Langevin stochastic process  \cite{FELL95}
implemented using the simplectic algorithm \cite{KOLB99}.  For the
mass $Q$ and the friction constant $\gamma_V$ of the piston  acting on
the simulation box, we use $Q = 0.054 / R_{ww}^4$ and  $\gamma_V =
0.5$. A parallelogram with equal sides and defined by an  angle of
$120^{\circ}$ is used for the simulation box. This geometry  retains
the periodicity of a crystal made of water molecules through  the
boundaries. Periodic boundary conditions are implemented using the
minimum image convention. For the Langevin equations describing the
motion of particles, we use the friction constant  $\gamma^{-1} = 0.93
\tau_o$. The noise  term in the Langevin equations of motion is given
by the  fluctuation-dissipation theorem. Pressure is set to 0.2 in
units  of $\epsilon_h/R_h^2$. At this pressure, the MB model
reproduces  water-like anomalies seen at ambient pressure
\cite{DILL05} and  hydrates non-polar molecules in a realistic manner
\cite{SOUT02}.  The simulation box is packed with 512 molecules
comprised of  a 10-monomer long protein and 502 water molecules. To
represent  the solvent in its liquid state, we use temperatures
ranging from  0.145 to 0.25 in units of $\epsilon_H$.

The system was initially equilibrated at a temperature of 0.25 for
5000 time steps, followed by a data collection period of 50000 time
steps. The temperature was then lowered and the
equilibration-collection  cycle was repeated. This cooling procedure
was repeated until the  lowest temperature was reached. Four samples
with different  initial conditions were prepared using this protocol
and the  distribution of the protein's radius of gyration $R_G$
\cite{doiedwards}  was computed. To obtain equilibrium properties, the
final configuration  at each temperature was used to extend the
simulation time until  the distribution of $R_G$ of the four samples
converged within  a root-mean-square value of 0.02.

\begin{figure}[tb]
\vspace{0.in} \epsfxsize=3.25in {\epsfbox{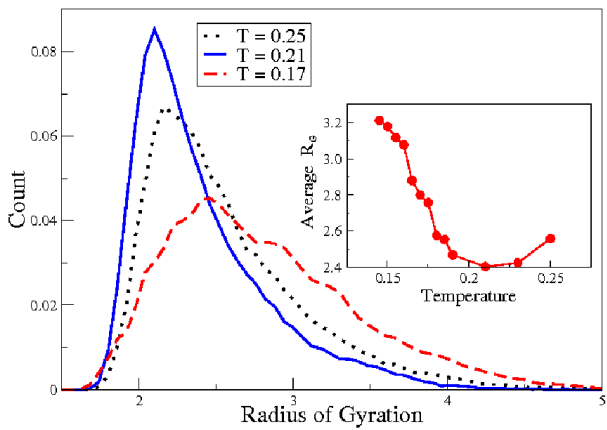}}
\caption{\label{fig:gyration} (Color Online)  Normalized distribution
of the radius of gyration $R_G$ at three  temperatures: $T = 0.25$, $T
= 0.21$ and $T = 0.17$.  Inset: The temperature dependence of $R_G$ of
the protein.}
\end{figure}

Let us now move on to describe the results.  In
Fig.~\ref{fig:gyration}, we show the equilibrium distribution  of
$R_G$ averaged over the four samples at three different
temperatures. An initial decrease in temperature, from 0.25 to 0.21,
shifts the peak of the distribution to a lower value. Therefore,  in
{\em hot water}, proteins favor more compact configurations  when the
temperature of the system is lowered. However, a further  decrease of
temperature results in completely different behavior:  as the
temperature decreases from 0.21 to 0.17, the peak shifts to  a larger
value indicating that {\em in cold water} proteins become  less
compact for decreasing temperature. This behavior in hot and  cold
water is shown systematically in the inset of
Figure~\ref{fig:gyration},  which depicts the temperature dependence
of $R_G$. The radius of  gyration decreases as temperature decreases
towards 0.21 -- in hot  water the protein folds as temperature
decreases. Below that temperature,  $R_G$ increases monotonically as
temperature decreases --  in cold  water the protein unfolds as
temperature decreases. These two types  of behavior are characteristic
of heat and cold denaturation of real  proteins and are in line with
previous studies \cite{KUNU02,RIOS00,PASC05}.

\begin{figure}[tb]
\begin{center}
\vspace{0.in} \epsfxsize=3.25in {\epsfbox{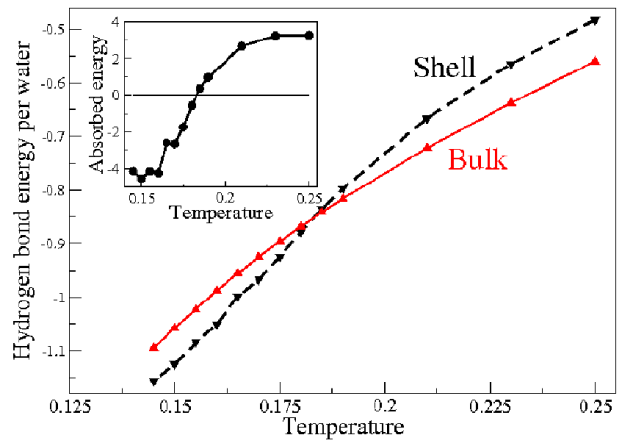}}
\caption{\label{fig:energy}  (Color Online)  Hydrogen bond energy per
water molecule for shell and bulk water.  Inset: Absorbed energy to
accommodate the protein at different  temperatures. The shell is
defined by water molecules whose  distance to the protein is less than
2.5 in units of $R_h$.}
\end{center}
\end{figure}

The parabolic-like shape of $R_G$ (see the inset of
Fig.~\ref{fig:gyration}) cannot be mapped into a model with local
monomer-monomer interactions only \cite{BRUS00}. To study the role  of
water, we show in Fig.~\ref{fig:energy} the average H-bond  energy per
water molecule for shell and bulk water.  The energy of shell water
averaged over the different  configurations is higher than the energy
of bulk water at high  temperatures. This changes gradually as
temperature decreases  such that the creation of shell water becomes
energetically  favorable at low temperatures. Therefore, when a
protein is  immersed in cold water it releases heat to form the shell,
while  in hot water it absorbs heat.  These features are again
characteristic of cold and heat denaturation  of real proteins
\cite{PRIV86}. In the inset of Fig.~\ref{fig:energy}  we show the
energy absorbed by the system to create the shell  around the
protein. The absorbed energy is defined as the difference  in H-bond
energy between shell and bulk water multiplied by the average  number
of molecules forming the shell. The absorbed energy decreases
monotonically with decreasing temperature and becomes negative below
some $T$ indicating heat release.

\begin{figure}[tb]
\hspace{0.0in} \epsfxsize=3.25in {\epsfbox{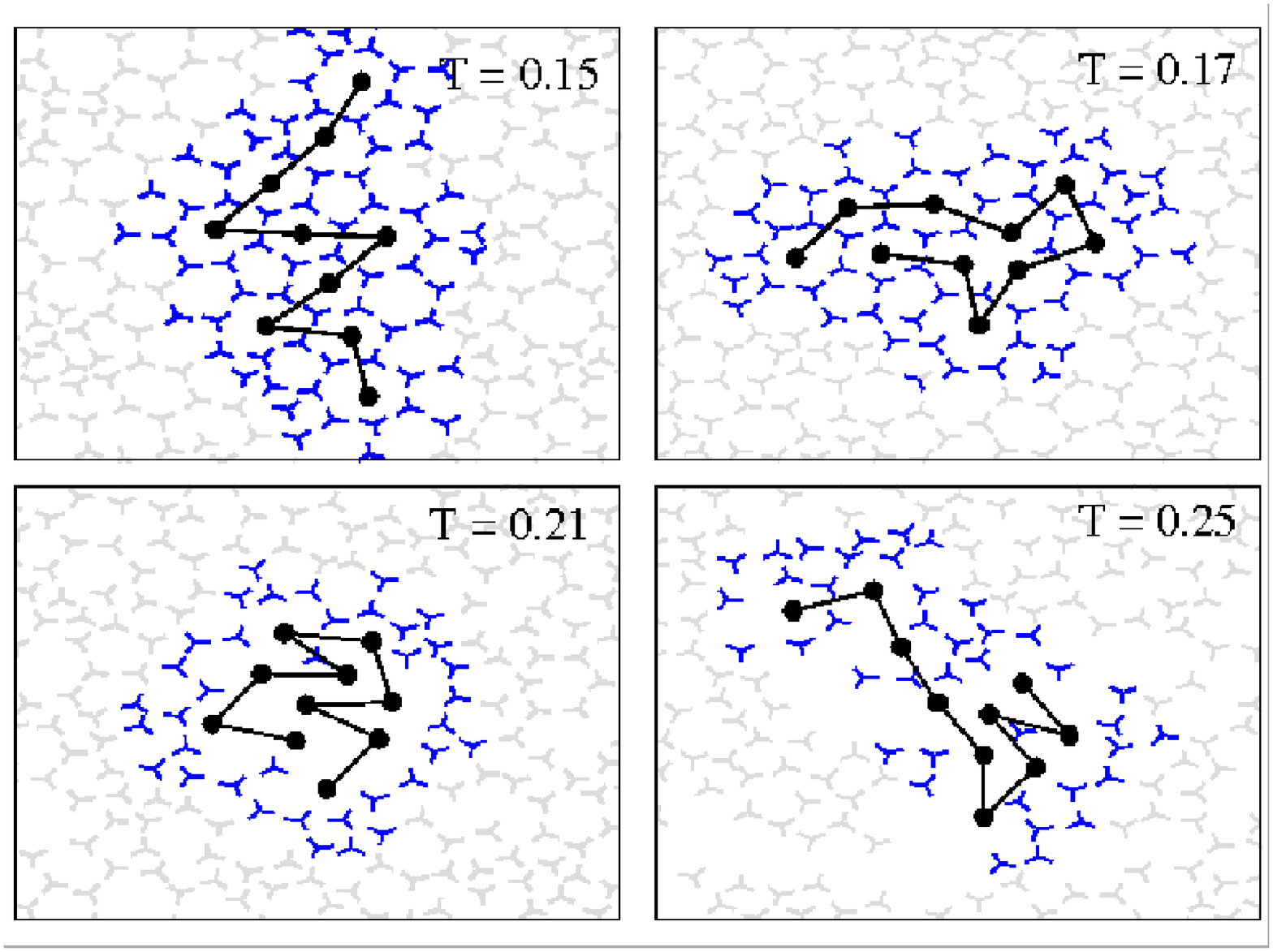}}
\caption{\label{fig:configuration} (Color Online)  Characteristic
configurations of a protein in cold water  ($T = 0.15$ and $T =
0.17$), at an intermediate temperature ($T = 0.21$), and in hot water
($T = 0.25$). The distance  of highlighted (shell) water molecules to
the protein is  less than 2.5 in units of $R_h$. In cold water, the
monomers  are typically surrounded by clathrate-like cages.}
\end{figure}

Characteristic configurations of the protein at different temperatures
are shown in Fig.~\ref{fig:configuration}. In  cold water (upper
panels), the solvent forms a cage around  each monomer of the protein,
i.e. monomers are surrounded by  an ordered layer of water
molecules. Molecules forming the  cage are strongly H-bonded to each
other and therefore have  a low energy. At $T = 0.21$, the protein
favors compact  configurations. Water molecules close to the protein
have at  least one non-saturated H-bond which is pointing towards the
protein. When the temperature is increased to $T = 0.25$,  most
monomers are in contact with the solvent. The solvent  forms either
incomplete cages around monomers, i.e. cages  which do not surround
monomers from all sides, or they correspond  to particles which are
weakly bonded to the other solvent  particles and are thus
energetic. The crossover behavior  of shell water shown in
Fig.~\ref{fig:energy} is therefore  characterized by the formation of
cage-like configurations at  low temperatures and the presence of
dangling H-bonds at high  temperatures.

Configurations where monomers are separated by an ordered  layer of
solvent molecules have also been shown to become  more stable, as
temperature decreases, in models for the  hydrophobic effect of
methane-like solutes \cite{PASC04,SOUT02}.  Solvent layers around
those monomer-pairs are highly ordered such  that their formation
decreases the entropy of the system  \cite{entropy}. Unfolding at low
temperatures is therefore  accompanied by a lowering in the entropy of
the {\it total\/}  system in accordance with experiments
\cite{PRIV86}, shell water  molecules becoming more ordered as the
protein becomes less  ordered. This mechanism explains the
counter-intuitive decrease  in entropy during cold denaturation. The
phenomenology is as  follows. When non-polar solutes are transferred
into water, the  system relaxes by ordering those solvent molecules
around the  solute. This ordering has an entropic cost which is
minimized  by clustering non-polar solutes together, as this decreases
the amount of surface around solutes. As the temperature  decreases
below a particular value, the system rebuilds the ordered layer of
solvent around non-polar solutes to saturate the dangling H-bonds left
on the surface of clustered solutes -- minimizing the
enthalpy. Although hydrophobicity is not  the only force responsible
for the stability of proteins, the  formation of a hydrophobic core
plays the dominant role.

In conclusion, we find that, at low temperatures, shell water 
forms hydrogen-bonds better than bulk water. Microscopically 
this correlates with the presence of solvent-separated-configurations 
which accounts for the unfolding of the protein at low temperatures. 
The existence of such low energetic states for shell water at low 
$T$ explains why cold denaturation proceeds with heat 
release as opposed to heat absorption seen during heat denaturation. 
Although here we studied cold denaturation in two dimensions, 
solvent-separated-configurations have also been shown to become 
more favorable as temperature decreases in a 3D model for the 
hydrophobic effect \cite{PASC04}. Therefore we 
expect that the results found in this work remain valid in 
3D systems. Our results further suggest that cold and pressure 
denaturation could be studied under a single framework: 
a transition towards solvent-separated-configurations 
\cite{HUMM98}.

This work was supported by the Natural Sciences and Engineering
Research Council of Canada, and {\it le Fonds Qu\'{e}b\'{e}cois 
de la recherche sur la nature et les technologies\/}. IV and T.A-N. 
wish to thank support from the Academy of Finland through its 
COMP Center of Excellence and TransPoly grants.


\begin{thebibliography}{31}
\bibitem{entropy}{According to table 1 of Southall and Dill
\cite{SOUT02}, the entropy term ($-T\Delta S$) of contact
configurations is $-0.55$ while the same term for solvent separated
configuration is 0.18 -- for two monomers embedded in 120 water
molecules at $T=0.21$. Therefore the entropy decreases when
hydrophobic contacts are destabilized in favor of solvent separated
configurations.}
\end{thebibliography}
\end{document}